\documentclass[aip,apm,reprint]{revtex4-2}
\usepackage{amsmath}
\usepackage{amssymb}
\usepackage{graphicx}
\usepackage{color,xcolor}
\usepackage{bm}%
\usepackage[colorlinks=true,linkcolor=blue]{hyperref}%

\begin{document}
\title{Ferroic orders in two-dimensional transition/rare-earth metal halides}
\author{Ming An}
\author{Shuai Dong}
\email{sdong@seu.edu.cn}
\affiliation{School of Physics, Southeast University, Nanjing 211189, China}
\date{\today}

\begin{abstract}
Since the discovery of graphene, two-dimensional materials with atomic level thickness have rapidly grown to be a prosperous field of physical science with interdisciplinary interests, for their fascinating properties and broad applications. Very recently, the experimental observation of ferromagnetism in Cr$_2$Ge$_2$Te$_6$ bilayer and CrI$_3$ monolayer opened a door to pursuit long-absent intrinsic magnetic orders in two-dimensional materials. Meanwhile, the ferroelectricity was also experimentally found in SnTe monolayer and CuInP$_2$S$_6$ few layers. The emergence of these ferroic orders in the two-dimensional limit not only brings new challenges to our physical knowledge, but also provides more functionalities for potential applications. Among various two-dimensional ferroic ordered materials, transition/rare-earth metal halides and their derivants are very common. In this Research Update, based on transition/rare-earth metal halides, the physics of various ferroic orders in two-dimensional will be illustrated. The potential applications based on their magnetic and polar properties will also be discussed.
\end{abstract}
\maketitle

\section{Introduction}
For rich low-dimensional physics and promising potential of applications, the research enthusiasm on two-dimensional (2D) materials have been aroused since the successful exfoliation of graphene in 2004.\cite{Novoselov:Science} Thanks to the extensive efforts devoted to the field of 2D materials, great progresses have been achieved in the past decade. For example, numerous single element monolayers have been discovered and synthesized, such as borophene, silicone, phosphorene, germannene, and antimonene. All these single-element monolayers are small gapped semiconductors but with relatively poor mechanical and electrical properties compared with graphene. Furthermore, monolayers with binary elements have also been found, such as hexagonal boron nitride (BN), molybdenum disulfide (MoS$_2$) and so on.

However, all these 2D pioneers are plain semiconductors or semi-metals, in which the charge is the dominant degree of freedom. Other functional properties like magnetism and polarity, originated from other degrees of freedom (e.g., spin, orbital, and lattice) are generally inactive or negligible in these systems, which limits their application regions. Although, many {\it ab initio} calculations and some delicate experiments claimed the magnetism in 2D materials induced by surface decorations, boundary edges, doping, defects, or biased electric field, these types of magnetism are somewhat extrinsic. Despite of tremendous efforts, these extrinsic magnetism induced by external factors are still hindered by their weak magnetization, low working temperature, weak robustness, and uncontrollability.

In contrast, 2D intrinsic magnetic materials with spontaneous long-range spin orders and homogeneous stable magnetization are obviously more appealing, which are more important for spintronic devices in the nanoscale and even angstrom-scale. However, limited by the Mermin-Wagner theorem,\cite{Mermin:Prl} for spin systems with continuous rotational symmetry, e.g., isotropic Heisenberg model or $X$-$Y$ model, long-range magnetic orders in the 2D limit can not be spontaneously stabilized at finite temperatures, due to the thermal perturbation. Fortunately, this restriction can be avoided when any single-axis anisotropy exists, e.g., an easy axis of magnetocrystalline anisotropy or external magnetic field. For example, for standard 2D Ising spins on the square lattice, its ferromagnetic Curie temperature ($T_{\rm C}$) is $\sim2.27J$,\cite{Onsager:Pr} where $J$ is the exchange between nearest-neighbor spins. In addition, it should be noted that the Mermin-Wagner theorem is derived under the conditions of thermodynamic limit, which works for systems with infinite size. For finite size systems, the magnetic orders can be established once the magnetic correlation length is over the system size. Thus, the limit of Mermin-Wagner theorem, will not be a principal barrier for 2D magnetic ordering.\cite{Cortie:AFM,Gong:Sci}

While for polar systems, discrete polar axes always exist naturally, without the continuous rotational symmetry, which is advantageous to avoid the restriction of Mermin-Wagner theorem in the 2D limit.\cite{Wu:Wcms} However, there is another obstructive factor in practice, i.e., the depolarization fields from surfaces and/or interfaces. The discontinuous polarization component perpendicular to the surfaces/interfaces can generate an electrostatic field, which is the main driving force of depolarization.\cite{Dawber:Rmp} In addition, the breaking crystalline symmetry and many dangling bonds may lead to polar surfaces/interfaces and serious surface/interfacial reconstructions, which can be another source of depolarization effect. Comparing with conventional three-dimensional (3D) crystals, these surface effects are much more serious in 2D sheets due to the super surface-to-volume ratio. Similar surface effects also exist in magnetic films, leading to the common ``dead layer" in ultra-thin films.\cite{Fong:Sci}

In principle, 2D magnetic/polar materials can be obtained by reducing the thickness of 3D crystals to the 2D limit, e.g., by growing ultra-thin films down to atomic level thickness using molecular bean epitaxy (MBE) or pulsed laser deposition.\cite{Fong:Sci,Ji:Nat} However, the open surfaces remain a big challenge for alloys and oxides whose original 3D bonding networks are broken at surfaces.

In this sense, the van der Waals (vdW) magnetic/polar materials are highly valuable for the physics community and material community. According to the experience from past studies on 3D magnetic and polar systems, transitional metal elements or rare-earth elements seem to be essential to be involved as cations. And there are more chances for halides than oxides to host these vdW magnets or polar systems, since halogen ions prefer less chemical bond coordinates than oxygen. Less chemical bonds are advantageous to form passivated surfaces of vdW sheets. Thus, transition metal or rare-earth metal halides are rich ore for 2D ferroic orders, which is the topic of this Research Update.

Thanks to the rapid developments of materials synthesis and in-suit measurements, remarkable progresses have been made in this field.\cite{Tang:JPCL} Due to the page limit and our knowledge limit, the aim of this Research Update is not going to be a comprehensive encyclopedia of all 2D magnetic and polar halides, but to clarify the main relevant physics based on selected examples since many materials share common physics.

This Research Update is organized as follow. The first section will describe the most representative Cr$X_3$ family, which have been extensively studied in recent years. The second section will review other 2D metal trihalide magnets, which share similar honeycomb lattice but much less verified. The third section will give a brief introduction to some metal dihalides, which own the triangular lattice and thus different physics. The geometric frustration may play a vital role in these systems. Last, several low-dimensional polar oxyhalides, with rectangular geometry, will be discussed. In some systems, both the magnetism and polarization exist and are coupled mutually.

\section{Chromium trihalides}
\subsection{Cr$X_3$ monolayer: 2D ferromagnets}
CrI$_3$ monolayer was one of the first two experimentally confirmed intrinsic 2D ferromagnetic materials,\cite{Huang:Nature} and another one was the Cr$_2$Ge$_2$Te$_6$ bilayer.\cite{Gong:Nature} Following this breakthrough, enormous efforts have been devoted to CrI$_3$, its derivatives, and its sister compounds. Prototype devices based CrI$_3$ monolayer or few layers have also been designed and fabricated to demonstrate the spintronic functions for applications.

\begin{figure}
\centering
\includegraphics[width=0.46\textwidth]{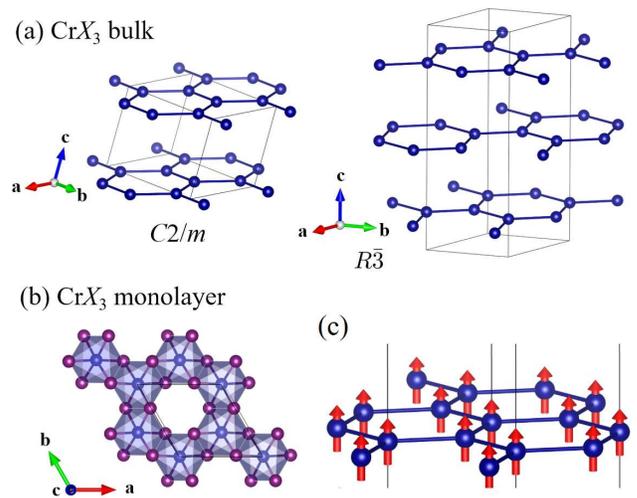}
\caption{(a) Schematic structure of Cr$X_3$ bulk. Left: the high temperature one; Right: the low temperature one. Only Cr ions are shown for simplify, which form the honeycomb structure. (b) Top view of Cr$X_3$ monolayer. (c) Side view of ferromagnetic spin order on Cr sites.}
\label{CrX3}
\end{figure}

Cr$X_3$ ($X$=Cl, Br, and I) bulks share the identical in-plane symmetry. As shown in Fig.~\ref{CrX3}(a-b), each Cr$^{3+}$ is caged in a halogen octahedron and neighboring octahedra connect in the edge-sharing manner, forming the hexagonal honeycomb lattice. Upon cooling, there is a structural transition from monoclinic ($C2/m$, AlCl$_3$ structure) to rhombohedral ($R\bar{3}$, BiI$_3$ structure) corresponding to the interlayer stacking sliding, which occurs at about $240$/$420$/$210-220$ K for $X$=Cl/Br/I respectively.\cite{Morosin:JCP,McGuire:CM}

\begin{figure*}
\centering
\includegraphics[width=0.98\textwidth]{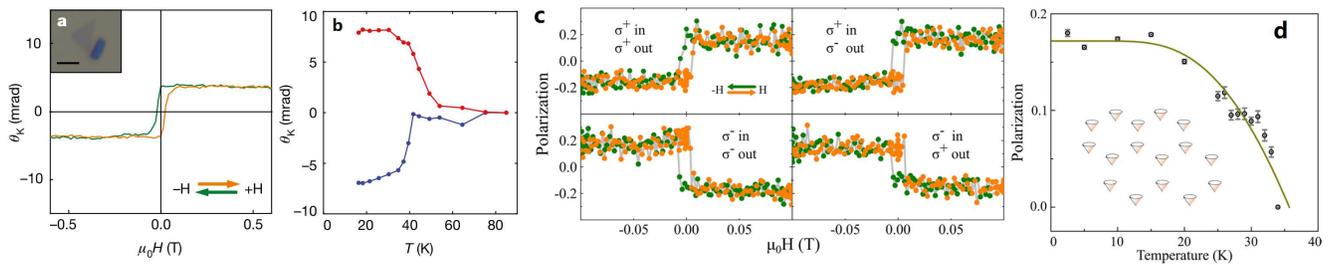}
\caption{(a) MOKE signal of CrI$_3$ monolayer at $15$ K as a function of magnetic field. The ferromagnetic hysteresis loop is observed. (b) MOKE signal of CrI$_3$ monolayer as a function of temperature. Blue: zero-field cooling. Red: field cooling at $0.15$ T. (c) Polarization resolved magneto-photoluminescence for CrBr$_3$ monolayer. $\sigma^+$ ($\sigma^-$) denotes the left (right) circularly polarized light. Ferromagnetic hysteresis loops are observed. (d) The signal of CrBr$_3$ monolayer as a function of temperature, suggesting a $T_{\rm C}$ at $\sim34$ K. (a-b) Reprinted with permission from Huang {\it et al.}, Nature (London) {\bf 546}, 270 (2017). Copyright 2017 Springer Nature. (c-d) Reprinted with permission from Zhang {\it et al.}, Nano Lett. {\bf 19}, 3138 (2019). Copyright 2019 American Chemical Society.}
\label{CrX3-mag}
\end{figure*}

Actually, as early as in the 1960s, unambiguous in-plane ferromagnetism [Fig.~\ref{CrX3}(c)] was confirmed in Cr$X_3$ crystals, whose magnetic transition temperatures are around $17$/$37$/$61(8)$ K for $X$=Cl/Br/I respectively.\cite{Dillon:JAP,Tsubokawa:JPSJ,McGuire:CM} This increasing trend can be attributed to two reasons:\cite{Zhang:JMCC} first, the net ferromagnetic exchanges (in fact a cooperative effect from various competitive exchanges) are weaken when the lattices shrink; second, the spin-orbit coupling increases with the atomic number, which enhances the magnetic anisotropy, an essential factor for long-range magnetic ordering in low-dimensions. Furthermore, a recent experimental study on CrI$_3$ crystal revealed a strong Kitaev interaction between spins, which is also responsible for higher $T_{\rm C}$.\cite{Lee:Prl20} The magnetic moments of Cr in all cases are close to the expected $3$ $\mu_{\rm B}$.\cite{McGuire:CM} The interlayer coupling is ferromagnetic for $X$=Br and I,\cite{McGuire:CM} but antiferromagnetic for $X$=Cl.\cite{Cable:PR} The magnetocrystalline anisotropy is also different between CrBr$_3$/CrI$_3$ and CrCl$_3$. In short, CrBr$_3$ and CrI$_3$ are similar layered ferromagnets with spins pointing out-of-plane, but CrCl$_3$ is a unique layered antiferromagnet with spins lying in plane.\cite{McGuire:CM,Cable:PR}

Based on the first principles calculations, the estimated cleavage energies indicate the easy preparation of monolayers for all Cr$X_3$'s.\cite{Zhang:JMCC,McGuire:CM} The stability of their free-standing monolayers was further confirmed by elastic property calculations and molecular dynamics simulations. The ferromagnetism was predicted to persevere down to monolayers for all Cr$X_3$'s, due to the robust intralayer ferromagnetic Cr-$X_2$-Cr super-exchange.\cite{Zhang:JMCC}

The experimental confirmation was delayed by technical difficulties in high-quality sample preparation and high-precision detection. Until 2017, Huang {\it et al} successfully obtained the CrI$_3$ monolayer by mechanical exfoliation.\cite{Huang:Nature} As illustrated in Fig.~\ref{CrX3-mag}(a), the spontaneous magnetization with out-of-plane orientation was observed in CrI$_3$ monolayer using the magneto-optical Kerr effect (MOKE) microscopy, which is a powerful tool to monitor the tiny ferromagnetism signal in 2D sheets. The $T_{\rm C}$ of CrI$_3$ monolayer is $45$ K [Fig.~\ref{CrX3-mag}(b)],\cite{Huang:Nature} lower than its bulk value.

Very recently, the intrinsic ferromagnetism in CrBr$_3$ monolayer was also evidenced [Fig.~\ref{CrX3-mag}(c)].\cite{Zhang:NL,Chen:Sci}  The measured $T_{\rm C}$ of CrBr$_3$ monolayer was $34$ K [Fig.~\ref{CrX3-mag}(d)],\cite{Zhang:NL} slightly lower than its bulk value. Stable CrCl$_3$ few layers and monolayer were also successfully cleaved from crystal, although their magnetism need further direct experimental verification.\cite{McGuire:PRM,Cai:NL}

Despite the similarity, CrBr$_3$ is more stable in air than CrI$_3$.\cite{Huang:Nature,Shcherbakov:NL} In fact, CrI$_3$ flakes are easy to decompose under ambient condition and have to be protected under inert atmosphere, while CrBr$_3$ samples have much longer lifetime in air and do not require special protection. Thus, CrBr$_3$ may provide a better platform to research 2D magnetism and their applications.

\subsection{Cr$X_3$ few layers: antiferromagnetism {\it vs} ferromagnetism}
More interestingly, the CrI$_3$ bilayer exhibits an unexpected antiferromagnetic interlayer coupling,\cite{Huang:Nature} although its bulk is ferromagnetic. This antiferromagnetic interaction across the vdW gap is naturally fragile, which can be tuned by strain and stacking mode between layers. The antiferromagnetic bilayer was argued to stack in the manner of high-temperature one (i.e., $C2/m$, AlCl$_3$ structure), while the low-temperature stacking manner remains ferromagnetic.\cite{Soriano:SSC,Jang:PRM,Jiang:PRB,Sivadas:NL}

\begin{figure*}
\centering
\includegraphics[width=0.98\textwidth]{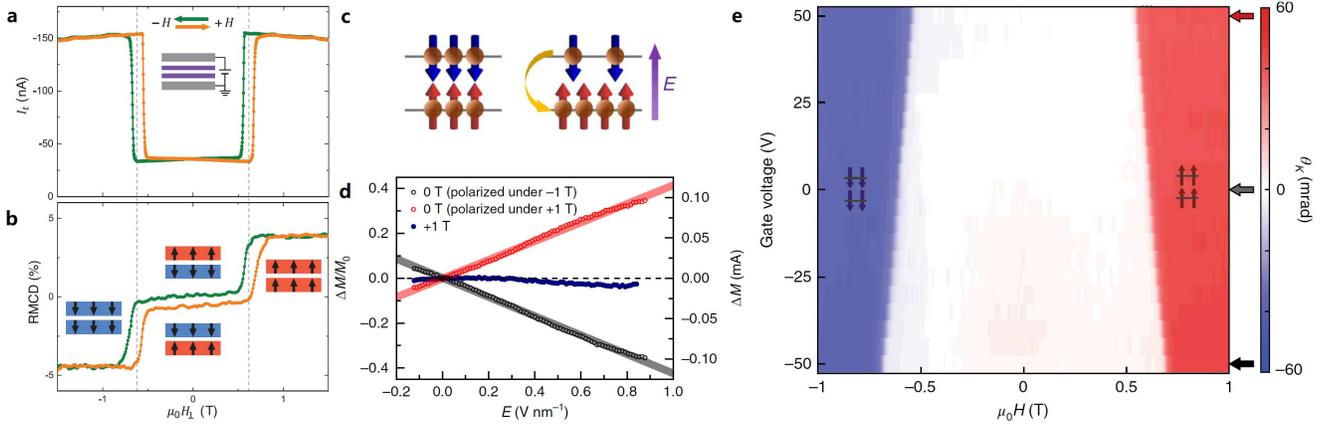}
\caption{Physical properties and functionalities of CrI$_3$ bilayer. (a) Giant tunneling magnetoresistance, which corresponds to (b) the meta-magnetic transition. (c-b) Linear converse magnetoelectric effect. (c) Schematic physical mechanism: the electric field can tune the charge distribution between two layers. (d) Experimental result of electric filed induced magnetization, which depends on the history of poling magnetic field. (e) Phase diagram characterized by MOKE signal as a function of electric and magnetic fields. Around the phase boundary between ferromagnetic and antiferromagnetic states, the electric field can tune the meta-magnetic phase transition. (a-b) Reprinted with permission from Song {\it et al.}, Science {\bf 360}, 1214 (2018). Copyright 2018 American Association for the Advancement of Science. (c-d) Reprinted with permission from Jiang {\it et al.}, Nat. Mater. {\bf 17}, 406 (2018). Copyright 2018 Springer Nature. (e) Reprinted with permission from Huang {\it et al.}, Nat. Nanotech. {\bf 13}, 544 (2018). Copyright 2018 Springer Nature.}
\label{CrI3bilayer}
\end{figure*}

When the layer number is further increased, the ferromagnetic state will be recovered and its $T_{\rm C}$ will restore to the bulk value,\cite{Huang:Nature} although a following work argued a ferrimagnetic state, i.e., antiferromagnetic coupling between layers.\cite{Song:Sci} This layer-dependent magnetic behavior greatly promoted the investigation of quantum magnetic phenomena and the development of spintronic devices.

In contrast, the interlayer coupling in CrBr$_3$ bilayer was determined to be ferromagnetic.\cite{Zhang:NL} However, the stacking-dependent interlayer coupling was also proposed for CrBr$_3$ bilayer.\cite{Chen:Sci} In details, using MBE, Chen {\it et al} successfully prepared the CrBr$_3$ monolayer and bilayer with two distinct stacking patterns (defined as H- and R-types). According to the hysteresis loops, H-type bilayer prefers ferromagnetic interlayer coupling, while in R-type sample the two layers are antiferromagnetically coupled. This phenomenon was somewhat similar (although not identical) to the situation of CrI$_3$ bilayer, providing a feasible new way to control magnetism in bilayers.

As a unique member, the layered antiferromagnetism in CrCl$_3$ can persist to bilayer,\cite{Cai:NL} while its stacking dependence has not been explored yet.

The antiferromagnetic CrI$_3$  bilayer can be easily driven to ferromagnetic via a meta-magnetic transition, with the critical magnetic field of about $0.6-0.8$ T, as depicted in Fig.~\ref{CrI3bilayer}(b) and (e).\cite{Huang:Nature,Jiang:NM,Huang:NN,Song:Sci}

Such layered antiferromagnetism can be implemented in magnetoresistance devices, which is of great importance in information processing and data storage. Indeed, giant tunneling magnetoresistance in vdW heterostructures has bee demonstrated with CrI$_3$ or CrCl$_3$ bilayers/trilayers acting as spin-filter barrier, as shown in Fig.~\ref{CrI3bilayer}(a-b).\cite{Song:Sci,Wang:Nc9,Cai:NL} And the related microscopic mechanism of spin-filtering has been clarified.\cite{Paudel:AMI} In turn, Klein {\it et al} utilized the tunnel conductance to probe the magnetic ground state and interlayer coupling.\cite{Klein:Science} Moreover, by combining the magnetoelectric effect and the spin filter effect, Jiang {\it et al} designed a spin tunnel field-effect transistor (FET) based on CrI$_3$ vdW heterostructure.\cite{Jiang:NE} Soon after, the on-off ratio of CrI$_3$ FET was successfully promoted to up to $10^4$.\cite{Patil:MTP} Additionally, many other interesting phenomena, such as magneto-optical effect, and spin-valley effect have also been investigated in CrI$_3$ sheets.\cite{Seyler:NP,Sun:Nature,Zhong:SA}

Besides these functionalities controlled by magnetic field, the converse magnetoelectric effect, i.e., to control magnetism using electric field, has also been demonstrated in CrI$_3$ bilayer, as shown in Fig.~\ref{CrI3bilayer}(c-e).\cite{Huang:NN,Jiang:NM} Two mechanisms were involved. First, upon the applied electric field, the electrostatic potential between two layers leads to uncompensated spin moments [Fig.~\ref{CrI3bilayer}(c-d)], i.e., a pure electrostatic effect for layered antiferromagnetism, and thus the linear magnetoelectric effect occurs. Second, near the vicinity of interlayer spin-flip transition, the CrI$_3$ bilayer can be switched between antiferromagnetism and ferromagnetism by electric field with the help of biased magnetic field, as shown in Fig.~\ref{CrI3bilayer}(e).

Such electric field induced meta-magnetic transition in CrI$_3$ bilayer was explained based on density functional theory (DFT) and tight-binding model calculations.\cite{Xu:JPCL} Later, Jiang {\it et al} revealed the electrostatic doping in CrI$_3$-graphene vertical heterostructures can tune the magnetism of both  CrI$_3$ monolayer and bilayer, even without the biased magnetic field.\cite{Jiang:NN} These results paved the way to pursuit voltage manipulation of spin state in the 2D limit.

\subsection{Modification of physical properties}
Besides these experimental attempts to utilize Cr$X_3$ monolayers and few layers, other modulations of Cr$X_3$ sheets have also been theoretically predicted. For example, to promote the ferromagnetic $T_{\rm C}$, Huang {\it et al} suggested an isovalent alloying method, which could enhance the $T_{\rm C}$ of CrWI$_6$ monolayer by $3-5$ times comparing with pristine CrI$_3$.\cite{Huang:JACS}

In addition, the strain and charge doping effects have also been revealed, exhibiting a rich magnetic phase diagram and evoking further experimental verifications.\cite{Zheng:Nanoscale} For example, phase transition from ferromagnetic state to antiferromagnetic one will occur when these trihalides monolayer are compressively strained.\cite{Webster:PRB} Contrarily, a moderate tensile strain will strengthen ferromagnetic coupling and enhance both magnetic anisotropy and transition temperatures of CrCl$_3$ and CrBr$_3$.\cite{Webster:PRB}

\begin{figure}
\centering
\includegraphics[width=0.46\textwidth]{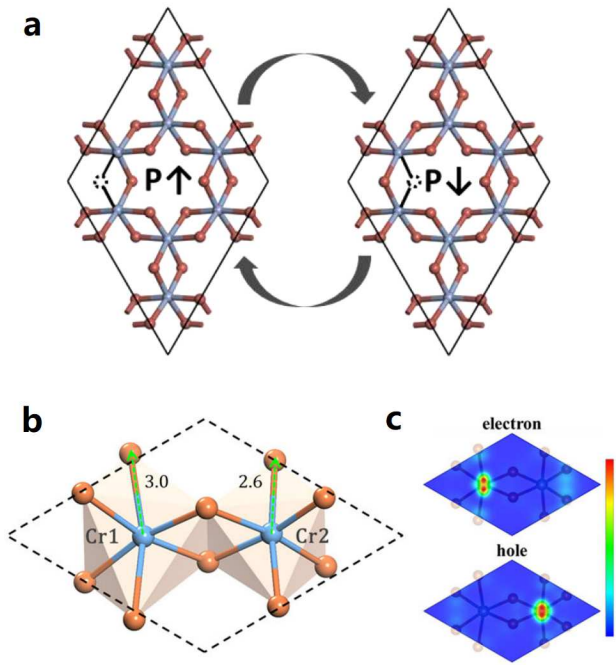}
\caption{Generating electric dipoles in Cr$X_3$ monolayers. (a) Out-of-plane polarization induced by an iodine vacancy. (b-c) Unequal Jahn-Teller distortion and charge density at neighbor Cr sites in half-doped CrBr$_3$ monolayer, which can generate in-plane polarization. (a) Reprinted with permission from Zhao {\it et al.}, Nano Lett. {\bf 18}, 2943 (2018). Copyright 2018 American Chemical Society. (b-c) Reprinted with permission from Huang {\it et al.}, Phys. Rev. Lett. {\bf 120}, 147601 (2018). Copyright 2018 American Physical Society.}
\label{CrX3-mod}
\end{figure}

Moreover, Zhao {\it et al} demonstrated that surface I vacancies of CrI$_3$ monolayer can enhance the local magnetic moment and increase $T_{\rm C}$.\cite{Zhao:NL} More intriguingly, the surface vacancy can induce an out-of-plane polarization, as sketched in Fig.~\ref{CrX3-mod}(a).\cite{Zhao:NL} The coexistence of stable ferromagnetism and switchable polarization make CrI$_{3-\delta}$ monolayer a magnetoelectric material. The point defect can even induce magnetic phase transition at certain concentration.\cite{Wang:CM}

Theoretically, Huang {\it et al} predicted 2D ferromagnetic ferroelectricity in charged CrBr$_3$ layer.\cite{Huang:PRL} Based on DFT calculations, they demonstrated that electron doping would induce unequal spatial electron-hole distribution on two Cr sites within CrBr$_3$ primitive cell [Fig.~\ref{CrX3-mod}(c)]. As a result of the charge ordering behavior, asymmetric Jahn-Teller distortion spontaneously takes place in two neighboring Cr-Br$_6$ octahedra [Fig.~\ref{CrX3-mod}(b)] and thus leads to orbital ordering. Due to the combination of charge ordering and orbital ordering, the pristine $D_{3d}$ symmetry is broken and replaced by the polar $C_2$ symmetry, meanwhile the in-plane polarization emerges. The possibility of magnetoelectric coupling was further proved by magnetization modulation with electric field.

All above theoretical predictions are valuable guides for experimental discoveries.

\section{Other $MX_3$: controversies to be verified}
Encouraged by the great progress of Cr$X_3$ family, it is natural to look at other $MX_3$'s, where $M$ can be V, Ni, Fe, Mn, Ru, etc.

Different from the widely studied Cr$X_3$, the studies on vanadium trihalides (V$X_3$) are quite rare. In 1980s, bulk VI$_3$ was first reported as a layered ferromagnetic insulator with $T_{\rm C}$ around $55$ K,\cite{Wilson:JPCSSP} which was confirmed by recent magnetization measurements ($50$ K).\cite{Tian:JACS} Its effective moment was about $2$ $\mu_{\rm B}$ per vanadium. Besides, the magnetic anisotropy was also observed with the easy axis normal to VI$_3$ sheets.\cite{Kong:AM,Son:PRB} The insulating band gap of $0.67$ eV was determined by electrical and optical transport measurements.\cite{Kong:AM}

Its structural space group was reported to be $R\bar{3}$ below $80$ K,\cite{Tian:JACS} identical to CrI$_3$. However, Son {\it et al} showed that its low temperature structure determined by refined powder X-ray diffraction was in better agreement with monoclinic $C2/c$ symmetry,\cite{Son:PRB} which was then confirmed by An {\it et al} through first-principles calculation.\cite{An:JPCC} Furthermore, Dole$\check{z}$al {\it et al} claimed another structural phase transition to triclinic upon further cooling down to $32$ K. Thus, further studies are necessary to clarify its ground state structure.\cite{Dolezal:PRM}

Despite the dispute of its bulk structure, the easy cleavable nature of VI$_3$ has been confirmed both theoretically and experimentally.\cite{Tian:JACS,Son:PRB,An:JPCC,He:JMCC} Based on DFT calculations, the ferromagnetic order can persist to monolayer.\cite{An:JPCC} The estimated $T_{\rm C}$ of monolayer is about $27$ K,\cite{An:JPCC} almost half of its bulk value. Such reduction of $T_{\rm C}$ was attributed to the reduced magnetic interaction due to the in-plane lattice expansion.\cite{An:JPCC} Some works claimed the change of magnetic anisotropy, i.e. from the easy axis off-plane in bulk to the easy $xy$ plane in monolayer,\cite{An:JPCC} while some others claimed unchange.\cite{Wang:PRB101,Yang:PRB101} More serious disagreement regarding the monolayer is that some work claimed a Dirac half metal,\cite{He:JMCC} while some predicted an insulator in agreement with its bulk.\cite{An:JPCC}

The close relationship between stacking pattern and interlayer magnetic coupling was also predicted in bilayer VI$_3$,\cite{Wang:PRB101} i.e., the ground state stacking order (AB stacking) prefers the ferromagnetic coupling, while the meta-stable one (AB' stacking) prefers the antiferromagnetic one.

As the sister members of VI$_3$, VCl$_3$ and VBr$_3$ are naturally expected to be promising 2D magnetic systems. Theoretical studies have been done to examine the magnetic properties of VCl$_3$ monolayer.\cite{He:JMCC,Zhou:SR6} Although the consistent conclusion of ferromagnetic ground state was reached, there was serious inconsistency regarding its $T_{\rm C}$, i.e., $80$ K predicted by He {\it et al} but over $400$ K by Zhou {\it et al}.\cite{He:JMCC,Zhou:SR6}

Further works, especially more precise experimental works, are highly desired to verify above inconsistent predictions on V$X_3$ sheets.

\begin{figure}
\centering
\includegraphics[width=0.46\textwidth]{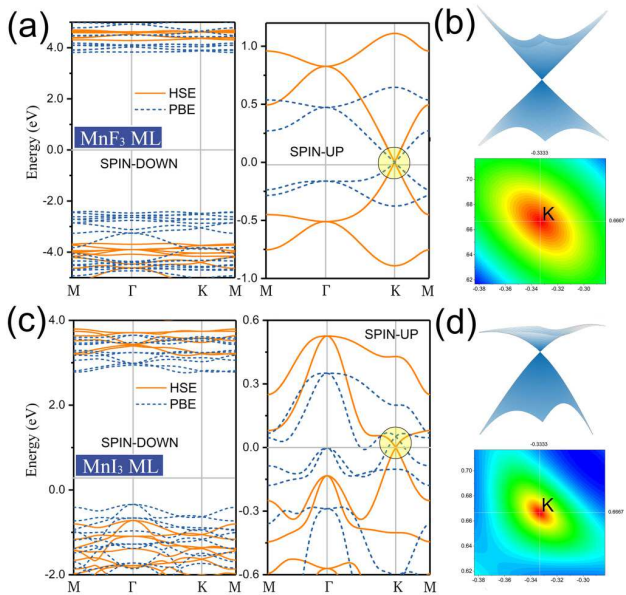}
\caption{Dirac half metal characteristic of Mn$X_3$ monolayers. (a) and (c) DFT Band structure of ferromagnetic MnF$_3$ and MnI$_3$ monolayer, respectively. (b) and (d) 3D band structure of the Dirac cone and its projection on the Brillouin zone. Reprinted with permission from Sun {\it et al.}, Phys. Rev. B {\bf 97}, 094408 (2018). Copyright 2018 American Physical Society.}
\label{MnI3}
\end{figure}

Besides V$X_3$, Mn$X_3$ is also a promising candidate family. Based on DFT calculations, Sun {\it et al} predicted that 2D stable hexagonal Mn$X_3$ monolayers possess large magnetic moment ($4$ $\mu_{\rm B}$/Mn),  in-plane magnetic anisotropy, high $T_{\rm C}$ above room temperature ($450-720$ K), and ideal Dirac half metal characteristic [Fig.~\ref{MnI3}].\cite{Sun:PRB97} Such high $T_{\rm C}$ must be overestimated considering the Ising model used. Similar to the V$X_3$, both the magnetic moment and the estimated transition temperature exhibit an increasing tendency with the atomic number of halogen ion in Mn$X_3$.\cite{Sun:PRB97}

Nickel trihalides were predicted to be another candidate family with ideal Dirac half metal nature and intrinsic 2D ferromagnetism above room temperature.\cite{Li:RSCA} The estimated magnetic moment was about $1$ $\mu_{\rm B}$ per Ni$^{3+}$, in consistent with its low spin configuration. Similar trend of magnetic enhancement was observed in Ni$X_3$ from Cl to I. The experimental feasibility of Ni$X_3$ monolayers was proved by stability analysis up to $500$ K.

For iron trihalides, the magnetism is not plain ferromagnetic. Early works reported the peculiar in-plane helimagnetic structure.\cite{Cable:PR,Johnson:JAP} Antiferromagnetic stacking was observed in FeCl$_3$ and FeBr$_3$ below $10$ K and $16$ K.\cite{Cable:PR,Johnson:JAP}

Besides these binary element $MX_3$, some combinations of $MM'X_6$ have also been calculated, which suggests more magnetic candidates to be tuned.\cite{An:JPCC}

The transition metal in $MX_3$ can also be heavier $4d$ ones, which own weaker Hubbard correlation but stronger spin-orbit coupling than $3d$ elements. More novel physical properties can be expected in these $4d$ trihalides. For example, the $\alpha$-RuCl$_3$ was supposed to be a model system for Kitaev interaction and a candidate of quantum spin liquid. However, complex low-temperature structure and multiple magnetic phase transitions were detected in $\alpha$-RuCl$_3$, which might be related to the symmetry difference caused by different stacking modes.\cite{Kobayashi:IC,Banerjee:NM} The only information available was its in-plane zig-zag antiferromagnetic state, with interlayer antiferromagnetic stacking order.\cite{Sears:Prb} The magnetic moments, including both spin and orbit components, were reported to have both in- and off-plane components, which further complicates the magnetic structure in $\alpha$-RuCl$_3$.\cite{Cao:PRB} The knowledge of magnetic nature of 2D $\alpha$-RuCl$_3$ monolayer or few layers remains very limited.

In short, other $MX_3$ families beyond Cr trihalides, can host exotic magnetism but the studies remain far from completed. Many theoretical predictions are full of uncertainness. Further careful calculations and experimental investigations are urgently needed.

\section{$MX_2$: triangular magnets}
Besides trihalides, metal dihalides are also available as vdW magnetic  materials. The key physics for these compounds is their 2D triangular lattice, which is geometrically frustrated for antiferromagnets.

The $MX_2$ bulks also have two types of stacking modes: the trigonal CdI$_2$ type (AA stacking) and the rhombohedral CdCl$_2$ one (ABC stacking). Structural transition between these two types are also possible upon the change of temperature and pressure. Within each layer, each $M$ ion is also caged in a $X$-octahedron and these octahedra connect with neighbors in the edge sharing mode.

\begin{figure}
\centering
\includegraphics[width=0.46\textwidth]{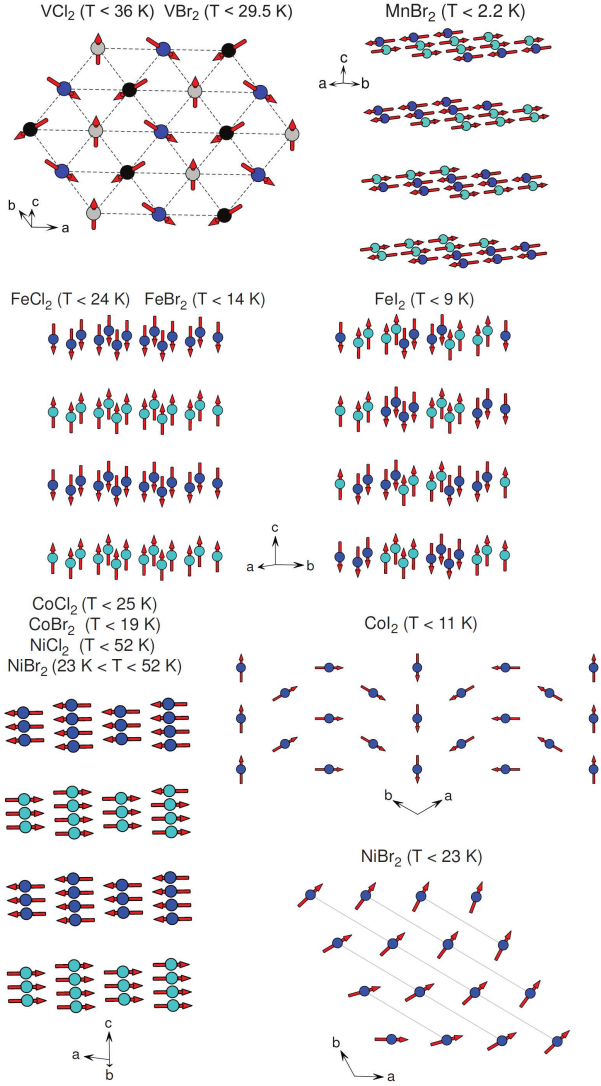}
\caption{Schematic of various magnetic orders in $MX_2$ bulks. The corresponding orders in monolayers of few layers have not been experimental verified but promisingly interesting. For those noncollinear spin orders, ferroelectric polarization may be generated. Reprinted with permission from McGuire, Crystals {\bf 7}, 121 (2017). Copyright 2017 McGuire.}
\label{MX2}
\end{figure}

Most known $MX_2$ bulks are antiferromagnetic, although some of them are ferromagnetic within each layer.\cite{McGuire:Crystal} Some typical magnetic textures are summarized in Fig.~\ref{MX2} by McGuire. The net exchanges within each layer may be complex, due to the multiple origins of exchanges. In particular, $MX_2$ ($M$=Fe, Co, $X$=Cl, Br) are ferromagnetic within each layer at low temperatures, but neighboring layers are antiferromagnetically coupled. However, their sister member FeI$_2$ is stripy antiferromagnetic in a layer, while CoI$_2$ owns the spiral antiferromagnetism as a ground state. More complexly, NiBr$_2$ is ferromagnetic in a layer in a middle temperature region (from $23$ K to $52$ K) but changes to a long period helimagnetic structure below $23$ K.
The magnetic transition temperatures of most $MX_2$ series decrease with the atom number of $X$, in opposite to the general tendency in most $MX_3$.\cite{McGuire:CM}

For $M$=V and Mn, antiferromagnetic coupling is expected within each layer, considering their $d^3$ and $d^5$ electron occupation (i.e., half-filling $t_{\rm 2g}$ orbitals or $3d$ orbitals). Then the geometrical frustration is unavoidable, which will reduce the ordering temperatures and generate noncollinear spin texture. The $120^\circ$ Y-type antiferromagnetism was reported for VCl$_2$ and VBr$_2$, as a typical result of geometrically frustration. More complicated helical magnetic structures were reported for VI$_2$, MnCl$_2$, and MnI$_2$, while MnBr$_2$ owns a stripy antiferromagnetism.

The noncollinear magnetic textures in these triangular lattices may break the spatial inversion symmetry and induce a ferroelectric polarization.\cite{Dong:Ap,Dong:NSR} In fact, for MnI$_2$, CoI$_2$, NiBr$_2$, and NiI$_2$ crystals, their magnetism driven ferroelectricity has been experimentally known for a long time,\cite{Kurumaji:Prl,Kurumaji:Prb}, which has also been explained.\cite{Xiang:Prl11} In fact, the relationship between $120^\circ$ Y-type antiferromagnetism and its ferroelectric polarization is a general physical phenomenon in 2D triangular magnets, including fluoridized MXene monolayer.\cite{Zhang:Jacs}

The monolayers of $MX_2$ have also been investigated theoretically. Fe$X_2$, Ni$X_2$ ($X$=Cl, Br, I), CoCl$_2$, and CoBr$_2$ monolayers were predicted to be ferromagnetic, while V$X_2$, Cr$X_2$, Mn$X_2$, and CoI$_2$ were predicted to be antiferromagnetic.\cite{Torun:Apl,Ashton:NL,Kulish:JMCC,Botana:Prm} It should be noted that in some of these theoretical studies, sometimes only simple antiferromagnetic textures were considered. Therefore, the real magnetic orders might be missed, and thus more careful studies are necessary. For example, a recent study predicted complex skyrmionic lattices in NiI$_2$ monolayer.\cite{Amoroso:arXiv}

Another interesting $MX_2$ system is GdI$_2$, which contains $4f$ magnetic moment and owns the MoS$_2$-type structure. The $4f$ electrons can contribute a larger magnetic moment up to $7$ $\mu_{\rm B}$ and relative larger magnetic anisotropy. The residual one $5d$ electron, playing as the intermediary, can align these $4f$ moments parallelly, persisting to the monolayer limit.\cite{Wang:MH}

The experimental investigations on $MX_2$ monolayers or few layers are not as plenty as $MX_3$, which call for more attentions in near future. Till now, only FeCl$_2$ and NiI$_2$ on substrates were reported,\cite{Zhou:Jpcc,Cai:Ns,Liu:AN} although their magnetism has not been characterized.

\section{Oxyhalides: rectangular lattice for polarity}
As  the counterpart of magnetism, polarity in solids are also highly valuable. In 2D materials, the first experimental discovered polar materials were SnTe monolayer with in-plane polarization and CuInP$_2$S$_6$ few layers with out-of-plane polarization.\cite{Chang:Science,Liu:Nc} Plenty exotic physical properties have been revealed, including the rare negative piezoelectricity.\cite{You:Sa} More details of novel physics and its potential applications of CuInP$_2$S$_6$ can be found in a recent review.\cite{Zhou:Fp}

As pointed in the beginning, the less bonding characteristic of halogens is helpful to obtain vdW gap and passivated surface. However, it seems that halogens do not prefer the polarity, at least for the proper polarity. Even though, in some vdW oxyhalides, the polarity can be obtained with the help of oxygen. Metal oxyhalides, in which anions include both oxygen and halogens, provide another low-dimensional platform with ferroic orders. These oxyhalides usually owns rectangular lattice, instead of honeycomb and triangular ones.

\begin{figure*}
\centering
\includegraphics[width=0.96\textwidth]{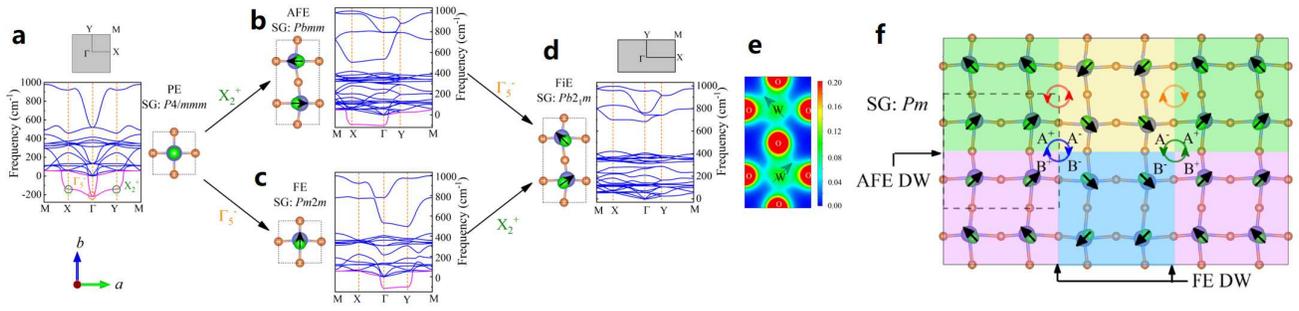}
\caption{Noncollinear ferrielectricity in WO$_2$Cl$_2$. (a-d) The phonon spectra of paraelectric (a), antiferroelectric (b), ferroelectric (e), and ferrielectric (d) states. (e) The $d^0$ rule driven polarity. (f) The $\mathbb{Z}_2\times\mathbb{Z}_2$ domain wall structures. Here the domains (A$^+$, A$^-$, B$^+$, B$^-$) are characterized by their ferroelectric and antiferroelectric phases. The atomic level dipole vortices/antivortices (red and orange circles) are formed at the ferroelectric domain boundary. The antiphase domain vortices/antivortices are denoted by blue and green circles. Reprinted with permission from Lin, {\it et al.} Phys. Rev. Lett. {\bf 123}, 067601 (2019). Copyright 2019 American Physical Society.}
\label{WOCl2}
\end{figure*}

The pristine $M$O$_3$ crystal owns a perovskite-like framework, in which $M$O$_6$ octahedra connect with each other via corner sharing oxygen ions. By replacing one divalent oxygen with two monovalent halogen ions, the $M$-O-$M$ chemical bonds between two adjacent layers will be cut off and replaced by a vdW gap. Based on this “scissors” effect, the three-dimensional $M$O$_3$ can be cut into 2D $M$O$_2X_2$ sheets,\cite{Lin:PRL} and even one-dimensional (1D) $M$O$X_4$ chains,\cite{Lin:PRM3} or even to zero-dimensional (0D) $MX_6$ molecules.

WO$_2$Cl$_2$ and MoO$_2$Br$_2$ monolayers are representative materials of these 2D dioxydihalide family. Noncollinear polarity was predicted by Lin {\it et al} through phonon spectrum analysis. As shown in Fig.~\ref{WOCl2}(a-c), there are two polar phonon modes: one ferroelectric and one antiferroelectric. The combination of these two modes leads to the novel noncollinear ferrielectric texture, as shown in Fig.~\ref{WOCl2}(d-e).

The polar distortions are driven by the $d^0$ rule and could hopefully strive up to room temperature.\cite{Lin:PRL} And more interestingly, such exotic noncollinear ferrielectric order can lead to novel physics such as unique dipole vortex at atomic level, topological $\mathbb{Z}_2\times\mathbb{Z}_2$ domain antiphase vortex [Fig.~\ref{WOCl2}(f)], as well as negative piezoelectricity.\cite{Lin:PRL}

\begin{figure}
\centering
\includegraphics[width=0.35\textwidth]{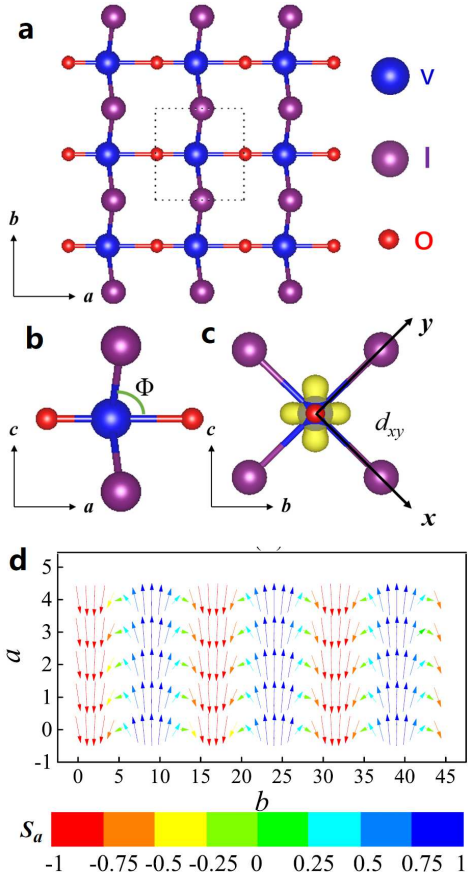}
\caption{Multiferroicity of VOI$_2$. (a-c) Schematic structure of VOI$_2$. (a) Top view. (b) Side view of a unit cell. (c) Side view from another side. The $d^1$ orbital shape is shown. (d) Possible magnetic ground state. A spiral spin order is induced by the strong Dzyaloshinskii-Moriya interaction associated with the polar distortion. Reprinted with permission from Ding, {\it et al.} Phys. Rev. B {\bf xx}, in press (2020). Copyright 2020 American Physical Society.}
\label{VOI2}
\end{figure}

Another concerned oxyhalide family is VO$X_2$ monolayer, as shown in Fig.~\ref{VOI2}(a-b). All VO$X_2$ monolayers were predicted to be ferroelectric and magnetic, i.e. multiferroics.\cite{Tan:PRB,Ai:Ns} In particular, the VOI$_2$ was predicted to be a rare ferromagnetic ferroelectric system.\cite{Tan:PRB} The ferroelectricity originates from the orbital ordering of $d^1$ configuration [Fig.~\ref{VOI2}(b-c)], which mimic the $d^0$ rule in the V-O-V bond direction.\cite{Ding:Prb} The magnetic moment from V$^{4+}$ is $1$ $\mu_{\rm B}$. However, due to the strong spin-orbit coupling of iodine, a strong Dzyaloshinskii-Moriya interaction is expected for the V-I$_2$-V bond, which is a side effect of ferroelectric polarization. Such Dzyaloshinskii-Moriya interaction will distort the magnetic texture to a spiral one [Fig.~\ref{VOI2}(d)], instead of collinear ferromagnetic state.\cite{Xu:Prl,Ding:Prb} Thus, the prediction of ferromagnetic ferroelectric ground state may be unavailable for VOI$_2$. Instead, another end member VOF$_2$ may provide the possibility to obtain ferromagnetic plus ferroelectric properties since its spin-orbit coupling is much weaker.\cite{You:Pccp}

In short, the oxyhalide families provide opportunities for both magnetism and polarity in 2D form. More experiments are encouraged in this direction.

\section{Summary \& perspective}
Obviously, the current research interests in low-dimensional functional materials is burgeoning. Here in this Research Update, the present status of some 2D vdW transition/rare-earth metal halides with intrinsic magnetism and polarity have been briefly summarized. Three selected categories have been presented: 1) $MX_3$ with honeycomb lattice; 2) $MX_2$ with triangular lattice; 3) $M$O$_2X_2$ and $M$O$X_2$ with rectangular lattice. Most of these systems are magnetic, some are polar, and a few of them are multiferroics. Plenty physical mechanisms are involved, including magnetic exchanges, geometric frustration, topological bands and domain structures, phonon instability, etc.

The theoretical efforts, most based on the DFT calculations, have predicted many interesting candidates and desired properties, while the experimental decisive investigations remain challenging due to the technical difficulties in the atomic scale. Even though, the theoretical calculations need to be more cautious, considering the subtle energies involved for vdW interactions, exchanges, anisotropy, dipole interactions, as well as spin-orbit coupling. All these interactions are ``weak", in the energy scale of meV or even $\mu$eV. Approximation and imprecision in calculations may lead to controversial results, like what happened in many $MX_3$'s except Cr$X_3$. In experiments, the substrate effects, imperfectness of samples, surface absorptions (i.e., atmosphere and humidity) and other extrinsic issues, may be much stronger in the energy scale and thus disturb the intrinsic ferroic properties.

Despite the great achievements in  the past years, there are more unknown issues or uncertainness in this emerging field. The following open questions are from our personal opinion.

Currently, the halogens involved in most of known 2D halides are Cl, Br, and I. However, the electronegativity of all these three elements are poor than oxygen. Thus, the chemical stability of these halides are mostly poor in air and against the humidity, which is not good for real applications. Oxygen and hydroxyl can replace these halogens easily, leading to the decomposition of these materials. This problem is most serious for those iodides. Then a natural expectation is whether fluorides can form these 2D halides? Theoretical calculations can be done easily for $M$F$_2$/$M$F$_3$/$M$O$_n$F$_2$. However, many corresponding fluorides are not vdW materials but 3D ionic crystals due to the very small size of fluorine ion. Are these 2D fluorides hopeless or promising?

In many 3D ionic crystals,  e.g., $AB$O$_3$ perovskites, there are two cation sites. The doping on A site can largely tune the electron density at B site but keep the structure unbroken. Thus the physical properties can be tuned continuously. However, in aforementioned halides, typically only one cation site is involved. The passivated surface of 2D halides and sole cation site make the electron density tuning at $M$ site difficult. Although electrostatic effect and surface absorptions/defects can do this job, they are not the robust routes for applications.  Then how to effectively tune the electron density in the cation site?

Most physical properties of 2D halides reviewed above inherit from their vdW bulks. In most cases, the ordering temperatures are reduced comparing with their corresponding bulks. Although some theoretical calculations predicted high ordering temperatures in some systems but none of them has been really confirmed experimentally yet. Is it possible for these 2D ferroic materials have improved performances, or even emergent new properties beyond bulks?

Most magnetic textures revealed in these 2D magnets are simple ones, although some noncollinear helical ones exist in these corresponding bulks or in DFT predictions. Recently, more complex skyrmions, have been predicted and demonstrated in the 2D limit,\cite{Liang:Prb20,Wu:NC20,Yuan:Prb20,Xu:Prb20,Amoroso:arXiv} which certainly deserves more careful studies.

Besides, the emergent physics of Moir\'e structures in twisted 2D bilayer have generated great enthusiasms in this field, especially for the twisted graphene.\cite{Cao:NAT18,Cao:Nature} It is natural to ask what will happen for the Moir\'e structures for 2D magnetic bilayers and polar bilayers?

Hopefully, this may be just the beginning of the era with rich physics to study and infinite opportunities to discover in the near further.

\begin{acknowledgments}
This work was supported by the National Natural Science Foundation of China (Grant No. 11834002).
\end{acknowledgments}

{\bf DATA AVAILABILITY}\\
Data sharing is not applicable to this article as no new data were created or analyzed in this
study.

\bibliographystyle{aipnum4-2}
\bibliography{ref3}
\end{document}